\documentclass{PoS} 

\newcommand{\beq}{\begin{equation}}
\newcommand{\eeq}{\end{equation}}
\newcommand{\beqa}{\begin{eqnarray}}
\newcommand{\eeqa}{\end{eqnarray}}
\newcommand{\om}{\Omega_m}
\newcommand{\lcdm}{$\Lambda$CDM} 

\title{A Cosmic Vision Beyond Einstein} 

\ShortTitle{Cosmic Vision Beyond Einstein} 

\author{\speaker{Eric V.\ Linder} \\ 
Berkeley Lab \& University of California, Berkeley USA} 

\abstract{
The acceleration of the cosmic expansion is a fundamental
challenge to standard models of particle physics and cosmology. 
The new physics of dark energy may lie in the nature of gravity, 
the quantum vacuum, or extra dimensions.  
I give a brief overview of the puzzles and possibilities of dark energy, 
and discuss the confrontation of a wide variety of ``beyond Einstein'' 
models with the latest 
data, showing what we currently know and what we must seek to learn. 
Next generation experiments using a variety of cosmological probes will 
deeply explore dark energy, dark matter, and gravitation. 
} 

\FullConference{Identification of Dark Matter 2008 \\ 
Stockholm, Sweden \\ 
August 18-22, 2008} 

\begin{document} 

\section{Introduction \label{sec:intro}}

What is the universe made of and how does it work?  These are the very 
basic, very fundamental questions displaying our current ignorance of 
95\% of the total energy in the universe.  But that we publicly ask 
them also indicates that we have hopes, and challenging plans, for how 
to answer them.  These questions and plans lie at the heart of the 
European Space Agency's Cosmic Vision program and the Joint Dark Energy 
Mission of NASA's Beyond Einstein program and the US Department of 
Energy.  

With the discovery of the accelerating expansion of the universe we 
clearly realize that the Standard Model of the particle physics of 
baryons, photons, etc., augmented by dark matter, is insufficient to 
explain the majority of the cosmic energy density.  The acceleration, 
or gravitational repulsion, property implies components beyond 
the standard ones dominated by rest mass or relativistic energy, to 
suggestions of quantum fluctuations of the vacuum, new high energy 
physics fields, or extensions of general relativity.  To 
see which way the physics lies, we indeed need a cosmic vision beyond 
Einstein. 

In \S\ref{sec:ideas} I present an overview of some of the key 
theoretical questions and the cosmological observational tools we 
have and can further develop to answer them.  Using the latest sets 
of data, \S\ref{sec:beylam} examines whether we have already narrowed 
down to a model essentially that of the static dark energy of 
Einstein's cosmological constant or whether there is still ample room 
beyond Einstein.  In \S\ref{sec:future}, methods for making our cosmic 
vision sharper, clearer, and steadier with space experiments are 
discussed. 

\section{Ideas about the Accelerating Universe \label{sec:ideas}} 

From the Equivalence Principle, we are used to thinking about the 
energy-momentum contents of the universe as inducing curvature in 
spacetime, the simplest picture being a ball of mass-energy bending 
the rubber sheet of spacetime such that a marble (test particle) 
bends its path toward it, experiencing an attraction we interpret 
as gravitation.  However, if the pressure contribution to the 
energy-momentum of a component 
is sufficiently negative, then the overall gravitational mass of 
the ball of ``negative pressure stuff'' -- given the more euphonious 
name of quintessence -- is itself 
negative \footnote{An alternative, not widely accepted \cite{ltb,eanna} 
idea for acceleration involves 
backreaction of nonlinear structure on the expansion.  This effectively 
seeks to use strongly positive pressure to density ratio (equation of 
state) but negative energy density.}. 
The rubber sheet potential well becomes a hill, and the marble retreats 
from the ball, experiencing repulsion.  This is one explanation of 
the cause of acceleration in the cosmic expansion: rather than the 
gravity of the average contents of the universe pulling things together 
and slowing the expansion, it pulls them apart, speeding up the cosmic 
expansion. 

The outstanding surprise about such repulsion is that it is not a highly 
exotic phenomenon witnessed only in extreme conditions in some small 
corner of the universe, but that it dominates our current universe, with 
over 70\% of the contents acting in such a mysterious manner.  One 
possibility for this physics beyond the standard model is Einstein's 
cosmological constant, an eternally unchanging pressure, negatively 
equal to its energy density, uniform in space and time.  Such constancy 
raises the question of why it comes to dominate the universe now, given 
our current picture of a universe of matter structures co-existing with 
acceleration: a factor four in the expansion factor into the past we 
would never 
have noticed observationally the cosmological constant, while a factor 
four in the future arrays of clusters of galaxies would be rare things to 
our telescopes.  See \cite{araa} for further discussion. 

Beyond the cosmological constant there are vast fields of dynamical 
(time varying) dark energy explanations, such as quintessence.  This 
dichotomy between static and dynamic is a fundamental issue to resolve. 
We need to know not just how much dark energy there is (its fractional 
contribution $\Omega_{\rm de}$ to the total energy density), but how 
springy/stretchy it is to the spacetime.  The ratio of the pressure to 
the energy density, called the equation of state $w$, is a (time 
dependent) measure of this.   By measuring through cosmological 
observations some instantaneous value $w$ and some measure of time 
variation $w'=dw/d\ln a$, where $a$ is the expansion factor (redshift 
$z=a^{-1}-1$), we can not only potentially exclude the cosmological 
constant ($w=-1$, $w'=0$) explanation, but actually be guided to a 
class of physics responsible for the explanation. 

\begin{figure} 
\begin{center}
\includegraphics[width=0.65\textwidth]{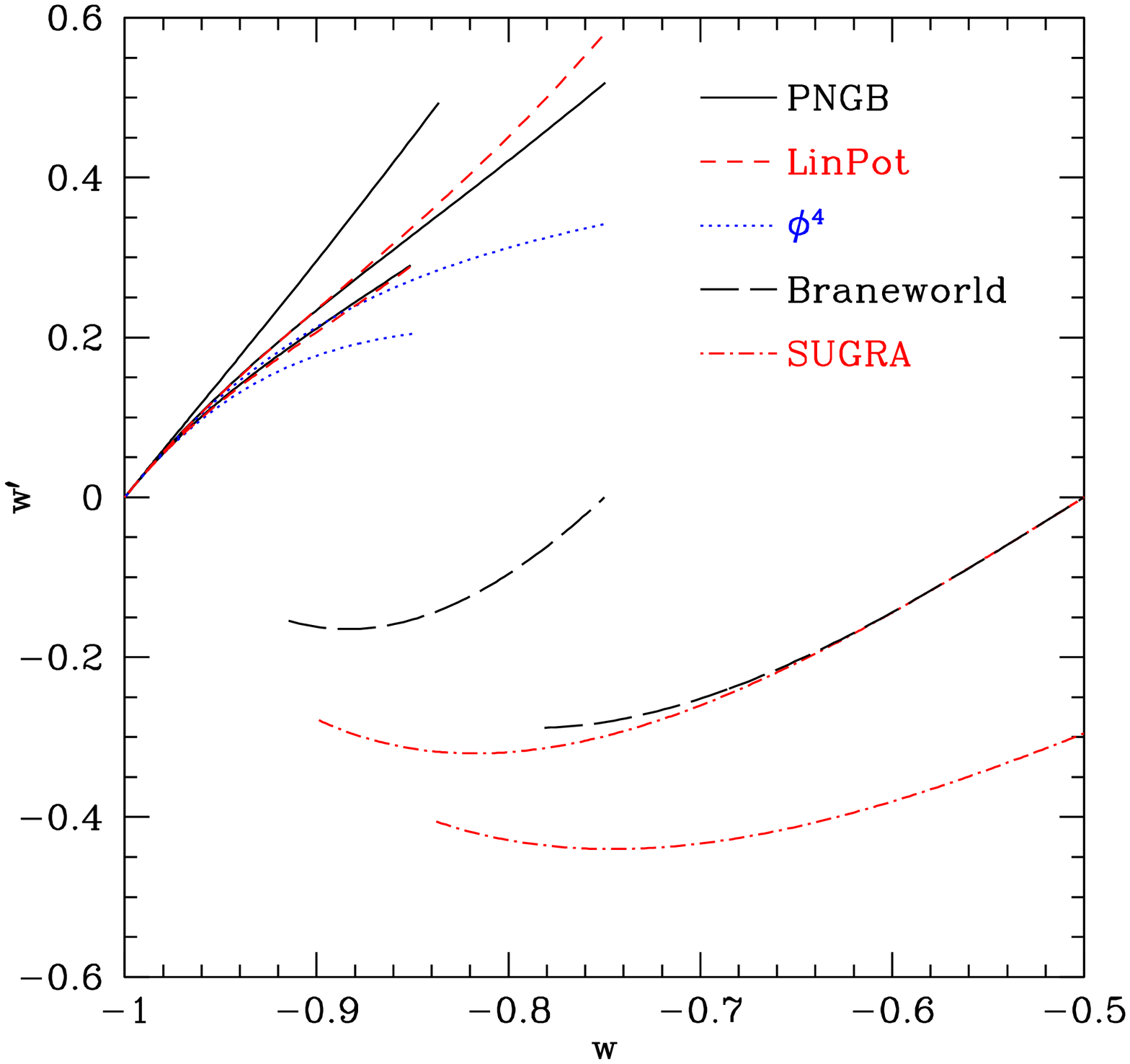} \\ 
\includegraphics[width=0.65\textwidth]{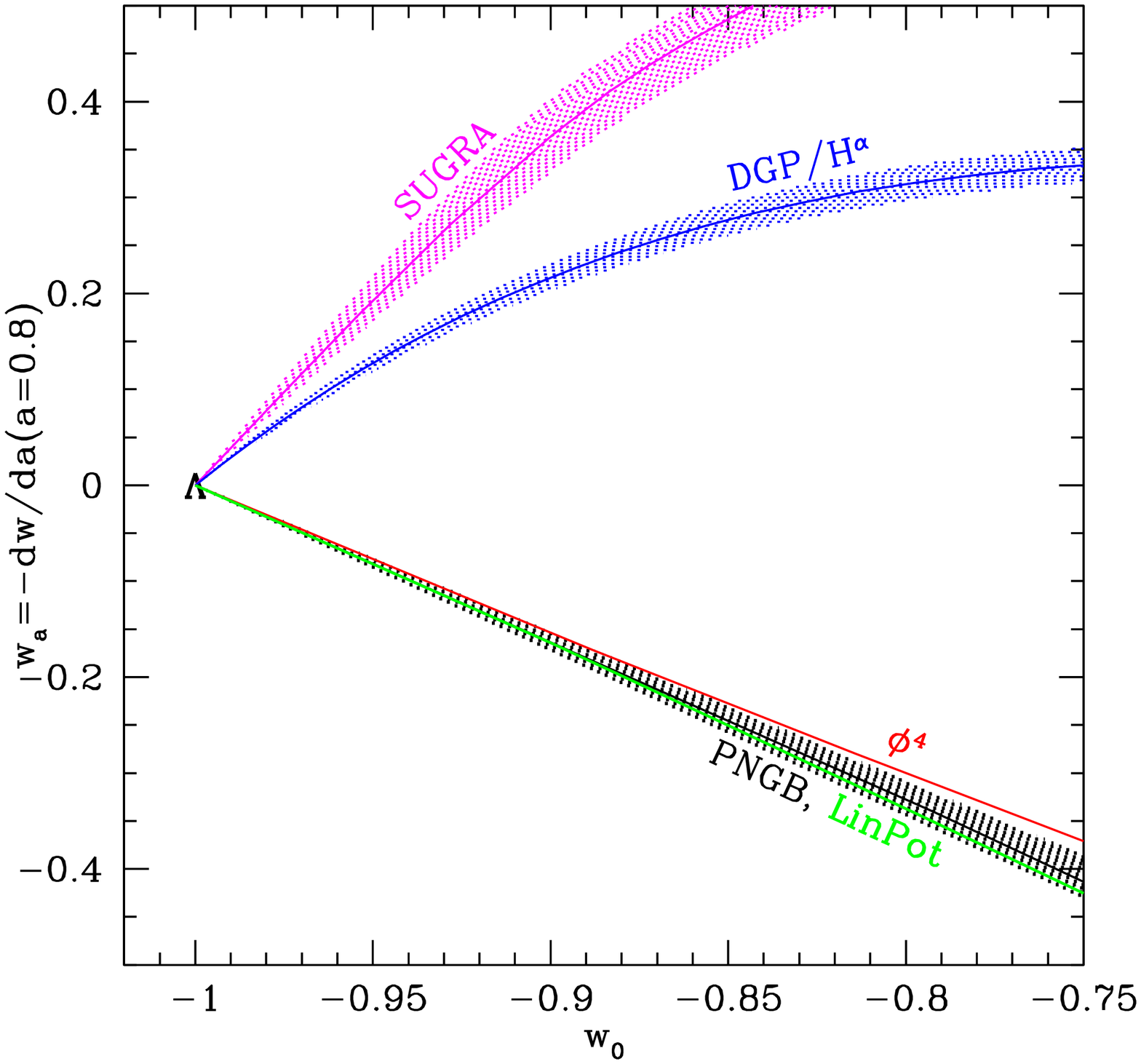}
\caption{The dynamics of dark energy models in the $w$-$w'$ phase space 
is diverse between models, and within models as the parameters of the 
potential are varied, although still separated into thawing (upper) and 
freezing (lower) regions [top panel].  In terms of the calibrated dark 
energy parameters $w_0$ and $w_a$, however [bottom panel; note vertical 
axis is flipped], models and 
families lie in tightly homogeneous regions, even scanning over the 
parameters within the potentials. From \cite{deplin}. 
} 
\label{fig:calib}
\end{center}
\end{figure}

\cite{caldlin} showed the utility of the $w$-$w'$ phase space to 
subdivide behaviors into thawing models (departing from a 
cosmological constant-like static behavior in the past) and freezing 
models (approaching cosmological constant behavior in the future). 
Just recently, \cite{deplin} demonstrated that proper choice of the 
particular measures of $w$ and $w'$ could highlight the differences 
in physics and separate dark energy models into distinct families 
(see Fig.~\ref{fig:calib}).  
Moreover, using these variables $w_0=w(a=1)$, the value today, and 
for the time variation $w_a=-dw/da(a=0.8)$, traces the observables of 
distances $d(z)$ and Hubble expansion rate $H(z)$ to 
$10^{-3}$ level accuracy -- better than needed for robust interpretation of 
next generation experiments.  Thus, a simple model-independent 
theoretical framework is in place for understanding what observations 
tell us about the accelerating universe. 

To carry out the mapping of the universe, we have several, complementary 
probes.  The cosmic microwave background radiation (CMB) is a direct 
probe of the early universe, and is highly useful for breaking degeneracies 
between dark energy and other parameters.  Three-dimensional surveys of 
galaxies and clusters of galaxies can study both expansion and the growth 
of structure, e.g.\ through the matter power spectrum and its baryon 
acoustic oscillation (BAO) features, and through weak gravitational lensing 
(WL).  Supernovae give a direct probe of the cosmic expansion over the entire 
acceleration epoch and back into the decelerating epoch.  These techniques 
have important complementarities, with the CMB and BAO data anchored in 
the high redshift, matter dominated era, WL data most sensitive at 
intermediate redshifts, and supernovae anchored in the current, accelerating 
era.  Together they give important crosschecks, systematics tests, and 
tight constraints on dark energy properties.

\section{Are We Done? \label{sec:beylam}} 

With such cosmological tools, it is not surprising there is a great deal 
of activity.  Unlike weather, where everybody talks about it but nobody 
does anything about it, with dark energy everybody wants to have something 
to do with it.  Most recently, the greatest observational advances 
have come from the Type Ia supernova technique.  \cite{kowal} carried 
out a complete reanalysis of all published data sets, amounting to 
307 supernovae after quality cuts, and applied robust, blind statistical 
and cosmological analysis.  Systematic uncertainties now contribute 
equally with statistical precision.  They find no significant deviations 
in the mean distance-redshift relation among the 13 data sets, no truly 
significant deviations in the redshift dependence or slope of the residuals 
from the mean, 
and no evidence of uncorrected evolution through tests of redshift 
and population subsamples.  Thus the cosmology results appear robust at 
the current level. 

Fitting for a constant $w$ equation of state by combining SN, BAO, and 
CMB data, \cite{kowal} finds $w=-0.969\pm0.061({\rm stat})\pm0.065 
({\rm sys})$.  This may lead some people to ask ``are we done?'', i.e.\ 
since $w$ is consistent with the cosmological constant value of $w=-1$, 
should we be content with that as the answer?  {\it Emphatically not\/}.  We 
do not know that $w(z)=-1$, allowing for time variation, or even anything 
at all about dark energy properties at $z>1$.  \cite{kowal} find 
constraints on the value 
of $w$ range from $-6$ to $+1$ at 68\% cl in a redshift bin from $z=1-2$. 

\begin{figure} 
\includegraphics[width=0.9\textwidth]{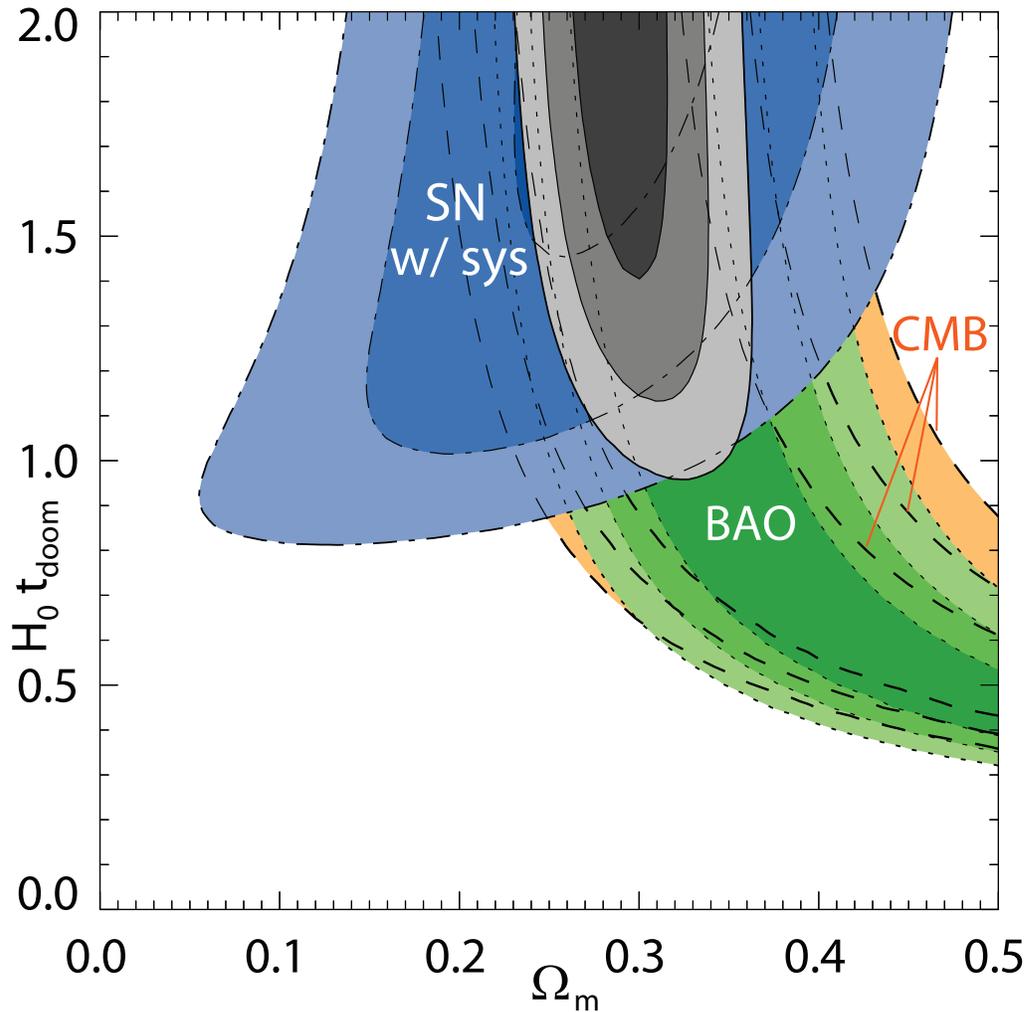} 
\caption{The future expansion history in the linear 
potential model has a collapse, or cosmic doomsday, at a finite time 
in the future.  While $t_{\rm doom}=\infty$ corresponds to the 
$\Lambda$CDM model, current data impose only $t_{\rm doom}>1.24 H_0^{-1}$ 
from now at 95\% confidence.  From \cite{rubin}. 
} 
\label{fig:lpdoom}
\end{figure}

While investigating redshift-bin values of $w$ or parametrizations gives 
great model-independent freedom, we can also compare specific physical 
models to the cosmological constant $\Lambda$ and ask whether the data 
prefer the 
cosmological constant.  This was done for 10 reasonably physically motivated 
models in \cite{rubin}, ranging over thawing (e.g.\ see 
Fig.~\ref{fig:lpdoom}) and freezing scalar fields, phase transitions, 
extended gravity, and geometric dark energy.  The results were that 
essentially all the models survive, with a couple in fact being modestly 
preferred over the cosmological constant.

The overriding conclusion is that essentially all types of physics are 
still in play, and that improvements in systematics control will have 
a major impact.  For example, the extra dimensional braneworld cosmology 
has a penalty of only $+2.7$ in $\Delta\chi^2$ relative to \lcdm, but 
this would increase to a highly significant $+15$ if the systematics 
uncertainty were removed.  A homogeneous data set of supernovae and 
a uniform analysis pipeline are key to revealing the physics, as will be 
the future sensitivity to the time variation $w(z)$.  
So to truly advance our knowledge of the nature of dark energy 
we must obtain better data, with next generation experiments.

\section{Clear Vision \label{sec:future}} 

\begin{figure} 
\includegraphics[width=0.9\textwidth]{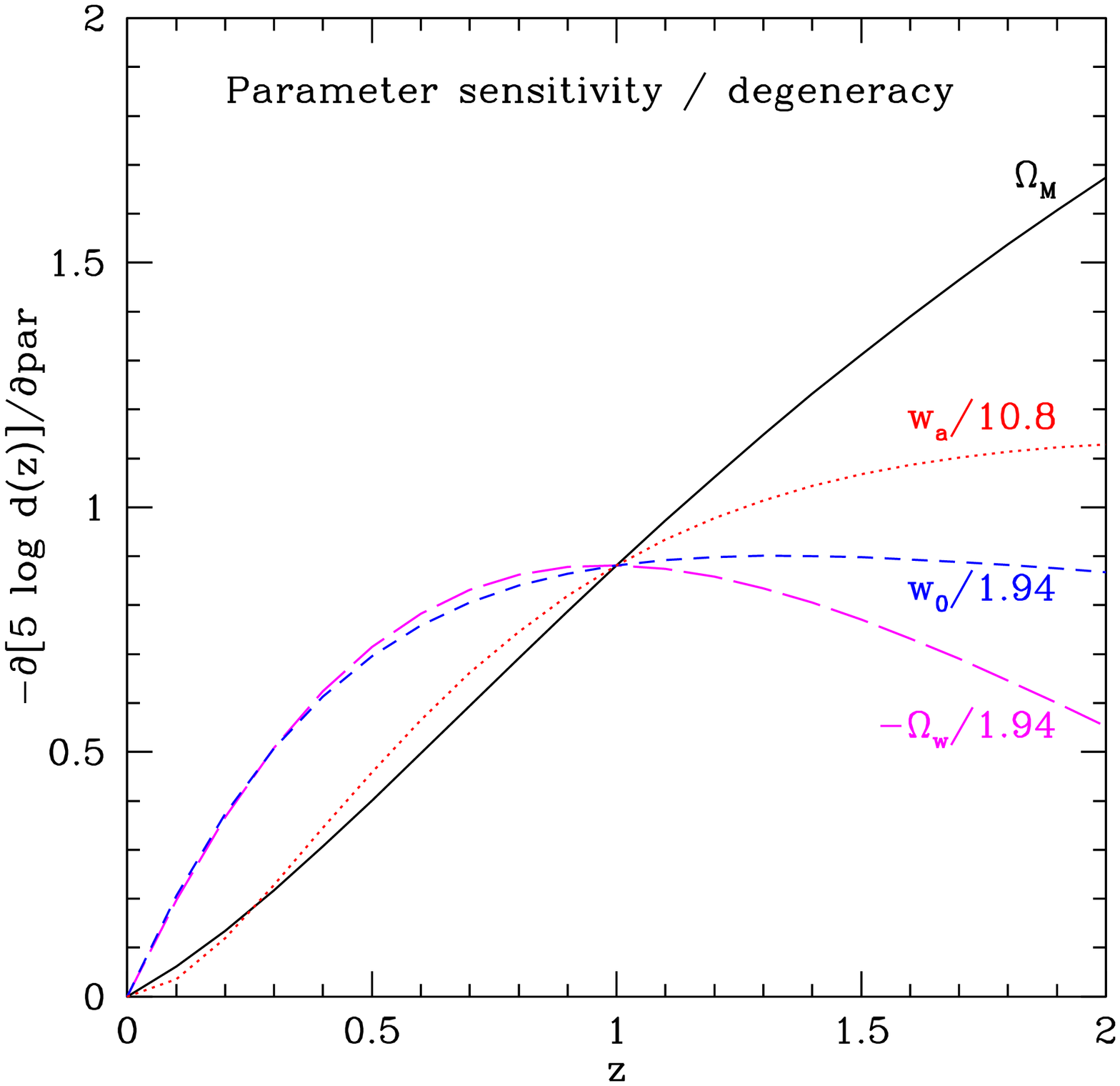} 
\caption{Cosmological parameter sensitivities for the magnitude-redshift 
relation, $5\log d(z)$, are plotted for the matter density $\om$, 
dark energy density $\Omega_w$ (spatial curvature is allowed), and 
dark energy equation of state parameters $w_0$ and $w_a$.  Since only 
the shapes of the curves matter for degeneracy, parameters are normalized 
to make this more evident.  Observations out to $z\ge1.5$ are required 
to break the degeneracies. From \cite{linrpp}. 
} 
\label{fig:space}
\end{figure}

To obtain the redshift range needed for sensitivity to the time variation 
$w(z)$ and to improve systematics control for the supernova and weak 
lensing methods, observations are driven to space.  Figure~\ref{fig:space} 
shows how the innate cosmology dependence and degeneracies between 
parameters requires high redshift observations.  

Space observations have strong advantages for both supernova and 
weak lensing probes.  For supernovae, the extended wavelength range 
from the optical into the near infrared allowed by being above the 
atmosphere gives a greater lever arm on dust extinction systematics, 
and the near infrared enables greater redshift depth, increasing 
cosmology sensitivity.  Weak lensing uses the apparent shape 
distortions of distant galaxies by intervening mass concentrations 
to probe combinations of distances and growth of structure.  The noise 
in these subtle (percent level) distortions is much less from the 
stable space environment than at any ground based telescope, which must 
contend with weather, gravity, and the atmosphere \cite{Kasliwal}, 
and the signal is clearer due to the high resolution of space 
observations and the stable point spread function \cite{WLspace}. 

Next generation space experiments such as from the Cosmic Vision program 
or Joint Dark Energy Mission will enable deep surveys mapping the last 
10 billion years of cosmic expansion over 10000 times the area of the Hubble 
Space Telescope Deep Field (HDF), and wide surveys of cosmic structure over 
10 million times 
the area of HDF, nearly as deep, and covering optical through near 
infrared wavelengths.  Widefield space surveys also maximize discovery 
potential, capable of mapping the dark matter skeleton of the universe over 
areas 5000 times that of the HST Cosmos survey \cite{Massey}, probing 
inflationary parameters and non-Gaussianity through precision measurements 
of the matter power spectrum, and assembling an impressive 40 trillion 
color-resolution elements on the sky -- 20 times that planned from next 
generation ground surveys.

\section{Summary \label{sec:concl}} 

Cosmic acceleration is a fundamental mystery demonstrating that our 
understanding of the universe and its contents is woefully incomplete. 
Indeed it puts us on the brink of a revolution in physics, though 
whether in quantum fields, gravitation, or cosmology is unknown.  Great 
efforts are being made to explore this, but current data do {\it not\/} 
tell us that the cosmological constant $\Lambda$ is the answer.  Indeed, 
two strikes against it are that Einstein and everyone else have failed 
for 90 years to explain the 120 orders of magnitude discrepancy from 
expectations, and that the one prior period of acceleration we know 
of -- inflation -- ended, and so clearly had time variation and dynamics. 

The need to understand the dynamics of the dark energy responsible for 
the current acceleration is crucial, and this requires going well beyond 
current data.  Fortunately, clear ideas exist of how to use several 
different, complementary probes in admittedly challenging, robust 
measurements.  With high resolution and low systematics, and multiple 
complementary probes of expansion and growth, we can robustly explore 
the next physics, testing cosmology, general relativity, quantum vacua, 
and dark matter.  Next generation space experiments will make sharper, 
clearer, and steadier our cosmic vision beyond Einstein.

\acknowledgments 

I am grateful to Joakim Edsj\"o and the organizers of IDM2008 for the 
invitation and wonderful hospitality.  
This work has been supported in part by the Director, Office of Science, 
Office of High Energy Physics, of the U.S.\ Department of Energy under 
Contract No.\ DE-AC02-05CH11231.

\end{document}